\documentclass[12pt,preprint]{aastex}

\newcommand{\kms}   {\mbox{{\rm km s$^{-1}$}}}

\newcommand{\um}{$\mu$m}
\newcommand{\hii}{\mbox{\ion{H}{2}}}

\newcommand{\degree}{^{\circ}}

\slugcomment{Accepted for publication in the Astrophysical Journal}

\shorttitle{Infrared Stellar-Wind Bowshocks}
\shortauthors{Povich et al.}

\begin{document}


\title{Interstellar Weather Vanes: GLIMPSE Mid-Infrared Stellar-Wind Bowshocks in M17 and RCW49}


\author{Matthew S. Povich\altaffilmark{1,5},
Robert A. Benjamin\altaffilmark{2,6}, 
Barbara A. Whitney\altaffilmark{3},
Brian L. Babler\altaffilmark{1}, 
R\'emy Indebetouw\altaffilmark{4}, 
Marilyn R. Meade\altaffilmark{1}, 
and Ed Churchwell\altaffilmark{1}}

\altaffiltext{1}{Dept.\ of Astronomy, University of Wisconsin-Madison, 475 N. Charter St., Madison, WI 53706}
\altaffiltext{2}{Dept.\ of Physics, University of Wisconsin-Whitewater, 800 W. Main St, Whitewater, WI 53190}
\altaffiltext{3}{Space Science Institute, 3100 Marine Street, Suite A353, Boulder, CO 80303-1058}
\altaffiltext{4}{Dept.\ of Astronomy, University of Virginia,
  Charlottesville, VA 22903-0818}
\altaffiltext{5}{email: povich@astro.wisc.edu}
\altaffiltext{6}{email: benjamin@wisp.physics.wisc.edu}




\begin{abstract}
We report the discovery of six infrared stellar-wind bowshocks in 
the Galactic massive star formation regions M17 and RCW49 from {\it Spitzer}
GLIMPSE (Galactic Legacy Infrared Mid-Plane 
Survey Extraordinaire) images.  
The InfraRed Array Camera (IRAC) on the {\it Spitzer Space
  Telescope} clearly resolves the arc-shaped emission 
produced by the bowshocks. We combine {\it Two Micron
  All-Sky Survey 
(2MASS),} {\it Spitzer,} {\it MSX,} and {\it IRAS} observations to obtain
the spectral energy distributions (SEDs) of the bowshocks and their
individual driving stars. We use the stellar SEDs to
estimate the spectral types of the three newly-identified O stars in
RCW49 and one previously undiscovered O star in M17.
One of the bowshocks in RCW49 reveals the presence of a large-scale
flow of gas escaping the \hii\ region at a few $10^2$ \kms. 
Radiation-transfer modeling of
the steep rise in the SED of this bowshock toward
longer mid-infrared wavelengths indicates that the emission is
coming principally from dust heated by the star driving the shock. 
The other 5 bowshocks occur where the stellar winds of O stars sweep
up dust in the expanding \hii\ regions.
\end{abstract}





\section{Introduction}

The Solar wind ends in a termination shock
\citep[e.g.][]{V105}, 
where the pressure of the heliosphere balances the ram pressure of the
surrounding 
interstellar medium (ISM). Massive stars with more energetic winds generate much
stronger shocks. 
In cases where
the relative
motion between the  
star driving the wind and the ambient ISM is large, the shock
will be bent back 
around the star. If the relative velocity is supersonic, the ambient
ISM gas is swept into a second shock, forming an arc-shaped
``bowshock'' that is separated from the termination shock by a contact
discontinuity. 
Stellar-wind bowshocks have been reported for a variety of sources, including
nearby runaway O stars \citep{vBu88,vBu95,Nor97,BB05,CP07,FML07}, high-mass
X-ray binaries \citep{EC92,K97,HK02}, LL Ori-type stars \citep{Bally00}, radio
pulsars \citep{GS06}, Galactic center O stars \citep{geb04,geb06}, and
mass-losing red giants \citep{Martin07}. Recently, an infrared (IR) bowshock has
been observed around the young A-type star $\delta$ Vel \citep{AG08}.
Cometary
\hii\ regions also resemble bowshocks, due either to density
gradients in the ambient gas or to motion of the ionizing source with respect to
the interstellar surroundings \citep{vB90,AH06}.  
Both the direction of a bowshock and its ``standoff distance'' from
the star generating the wind are determined by the velocity of the star with
respect to the surrounding medium. 
In the case of runaway O stars, this is
dominated by the high space motion of the star.

We report the
detection of three mid-IR bowshocks in each of two massive star formation
regions: M17 and RCW49. Two of the bowshocks in M17 are around known O
stars. We will demonstrate that the other bowshocks are also around likely O
stars. Since these stars are in or near expanding \hii\
regions, we find that 
the direction of the bowshock is determined principally by the flow of the ISM
rather than the space motion of the star.

\section{Observations and Interpretation} 

The Galactic Legacy Infrared Mid-Plane Survey Extraordinaire \citep[GLIMPSE;][]{ben03}
programs have mapped
the inner Galactic midplane ($|l| \le 65\degree$) using the IRAC
instrument on the {\it Spitzer Space Telescope}
\citep[2\arcsec\ resolution;][]{IRAC}. 
RCW49, located at $(l,b)=(284.3,-0.3)$, was observed as part of the validation of the
GLIMPSE observing strategy. 
The same area of the sky was
imaged 10 times with 2-second exposures,
and the combined data were mosaicked together
to produce high resolution images (0.6\arcsec\ pixels) of this region in the four IRAC bands: 3.6, 4.5,
5.8, and 8.0 \um. An overview
of these observations was given by \citet{chu04}. M17, at
$(l,b)=(15.0,-0.7)$, was included in the GLIMPSE survey area, with two 
visits on the sky combined to make mosaics with 0.6\arcsec\ pixels. 





\subsection{M17}

\begin{figure}[t]
\epsscale{.80}
\plotone{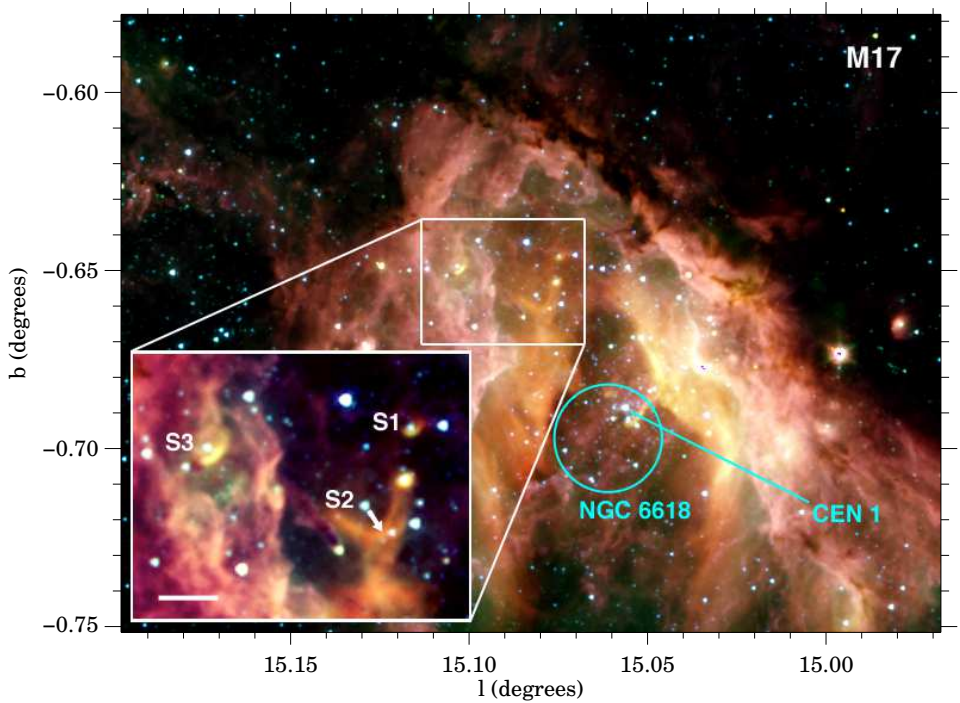}
\caption{GLIMPSE full-color image of M17 ({\it blue:} 3.6 \um, {\it
    green:} 4.5 \um, {\it orange:} 5.8 \um, {\it red:} 8.0 \um). The region
  containing the bowshocks  M17-S1, -S2, and -S3 is enlarged (scalebar
  shows $30\arcsec = 0.23$ pc at 1.6 kpc).  The central ring of O stars in the
  NGC 6618 cluster is circled. The bowshocks, along with the prominent
  pillar structure near  M17-S2, all appear to be oriented in the
  general direction of CEN 1, the O4-O4 binary system  
  in the center of the ionizing cluster.
\label{M17}}
\end{figure}
\citet{MSP07} studied the diffuse emission morphology of M17
at multiple wavelengths from IR to radio, %
constraining the distance to 1.4--1.9 kpc. We will assume the
widely-adopted value of 1.6 kpc for the M17 distance, following
\citet{MN01}. The 
winds and radiation of the 
O stars in
the NGC 6618 cluster have excavated a cavity in the center of the M17
\hii\ region. 
The cavity is filled with hot, X-ray-emitting gas
from shocked stellar winds \citep{T03}. \citet{MSP07} noted
the presence of 3 apparent 
stellar-wind bowshocks along the northern wall of the cavity, along
with a prominent ``elephant trunk,'' or pillar, all
oriented in the direction of the central ring of 7 O stars in NGC
6618. 
These structures are highlighted in Figure
\ref{M17}, a GLIMPSE image of M17 with an enlargement of the region
containing the bowshocks. The bowshocks stand out as yellow-orange
features in the image because they are faint at 3.6 \um\ (colored {\it
  blue}) compared to the other 3 IRAC bands.

We designate these stellar-wind bowshocks in Figure \ref{M17} with the
name of the 
region followed by an identification number, in order
of increasing Galactic longitude (for example, M17-S1). IR fluxes for all of the bowshock driving
stars are given in Table 1, and background-subtracted IR fluxes from
aperture photometry 
of the 
bowshocks are listed in Table 2.
\begin{deluxetable}{crrcccccc}
\tablecaption{2MASS and IRAC\tablenotemark{a} IR Fluxes for Bowshock Driving Stars (mJy)\label{tbl-1}}
\tablehead{
\colhead{ID \#} & \colhead{$l$} & \colhead{$b$} 
 & \colhead{$F[J]$}
 & \colhead{$F[H]$}
 & \colhead{$F[K]$}
 & \colhead{$F[3.6]$}
 & \colhead{$F[4.5]$\tablenotemark{b}}
 & \colhead{$F[5.8]$\tablenotemark{b}}
}
\startdata
M17-S1 & 15.07486 &  -0.64607 & 196 &  214 & 186 & 96 &  67 & \nodata
 \\
M17-S2 & 15.08126 & -0.65699 &  427 & 534 & 495 & 257 &  181 & 120  \\
M17-S3\tablenotemark{c} & 15.10325 &  -0.64867 & 103 & 200 & 238 & 151
& $\leq 154$ & $\leq 155$  \\
\tableline
RCW49-S1   &  284.07646 &  -0.43228 &  24 &  40 &  46 &  28 &  21 &  15 \\
RCW49-S2   &  284.30107 &  -0.37121 &  10 &  18 &  20 &  13 &
$\leq{11}$ &  $\leq 27$ \\
RCW49-S3   &  284.33999 &  -0.28269 &  83 &  97 &  90 &  47 & 32 &  27 \\
\enddata


\tablenotetext{a}{None of these stars was detected in the IRAC [8.0] band.}
\tablenotetext{b}{In cases where bowshock emission appears to
  cause a mid-IR excess over the stellar spectrum, 
  the [4.5] and [5.8] fluxes are
  reported as upper limits.
Due to suspected contamination from the bowshock or other diffuse emission, these 
  fluxes are treated as upper limits to the stellar flux.}
\tablenotetext{c}{Because the star driving SWB M17-3 is surrounded by bright,
  complex diffuse background emission, it was not extracted as part of
the GLIMPSE point-source catalog. The fluxes reported here were
measured using aperture photometry.}

\end{deluxetable}
\begin{deluxetable}{crrcccc}
\tablecaption{IR Fluxes for Bowshocks (mJy)\label{tbl-2}}
\tablewidth{0pt}
\tablehead{\multicolumn{3}{c}{} &
\multicolumn{4}{c}{IRAC Fluxes\tablenotemark{a}} \\
\colhead{ID \#} & \colhead{$l_{\rm apex}$} & \colhead{$b_{\rm apex}$} & \colhead{$F[3.6]$\tablenotemark{b}}
 & \colhead{$F[4.5]$}
 & \colhead{$F[5.8]$}
 & \colhead{$F[8.0]$}
}
\startdata
  M17-S1   & 15.0744 &  -0.6465 &  $\leq 34$  &  $71\pm 9$ &  $252\pm 42$ &  $1110\pm  290$ \\
  M17-S2   & 15.0791 &  -0.6613 &  $43\pm 17$ &  $245\pm
  48$ &  $1240\pm  159$ &  $8700\pm  1500$ \\
  M17-S3   & 15.1026 &  -0.6503 &  $69\pm   16$ &  $240\pm
  19$ &  $620\pm  100$ &  $1800\pm  320$ \\
\tableline
RCW49-S1 &  284.0775 &  -0.4305 & $5.7\pm 0.9$ &  $12\pm  1$ &  $33\pm
1$ &  $322\pm 5$ \\
RCW49-S2 &  284.3018 &  -0.3712 & $\leq 2$ &  $9.6\pm  0.5$ & $32\pm
3$ &  $172\pm  7$ \\
RCW49-S3 &  284.3388  & -0.2829 &  $\leq 21$ &  $50\pm  7$ & $160\pm
20$ & $825\pm 120$ \\
\tableline \tableline
\multicolumn{3}{c}{} &
\multicolumn{4}{c}{{\it MSX} Fluxes\tablenotemark{c}} \\
\colhead{} & \multicolumn{2}{c}{MSXC6 Name} & \colhead{$F[8.3]$} &
\colhead{$F[12.1]$} & \colhead{$F[14.6]$} & \colhead{$F[21.3]$} \\
\tableline
RCW49-S1 & \multicolumn{2}{c}{G284.0776-00.4340} & $627$ &
$2729$ & $5674$ & $1.56\times 10^4$ \\
\tableline \tableline
\multicolumn{3}{c}{} &
\multicolumn{4}{c}{{\it IRAS} Fluxes\tablenotemark{c}} \\
\colhead{} & \multicolumn{2}{c}{{\it IRAS} Name} & \colhead{$F[12]$} &
\colhead{$F[25]$} & \colhead{$F[60]$} & \colhead{$F[100]$} \\
\tableline
RCW49-S1 & \multicolumn{2}{c}{IRAS 10205-5729} & $3810$ &
$4.21\times 10^4$ & $\leq 2.26\times 10^5$ & $2.75\times 10^5$ \\ 
\enddata

\tablenotetext{a}{The IRAC fluxes are background-subtracted. Fluxes were measured using irregular apertures drawn to enclose all of the visible bowshock structure. Separate apertures were used to estimate the background flux. For each bowshock, the same set of apertures was used for all IRAC wavelengths.}
\tablenotetext{b}{In cases where stellar emission appears to be
  confused with bowshock emission, the [3.6] bowshock fluxes reported
  are upper limits only.}
\tablenotetext{c}{RCW49-S1 is a point source in both the {\it MSX} and {\it IRAS} catalogs. All of the other bowshocks are confused with bright diffuse background IR emission features at the resolutions of {\it MSX} and {\it IRAS}. 
}

\end{deluxetable}
Spectral types of two of the driving
stars have been determined photometrically \citep{Bum92} and spectroscopically
\citep{HHC97}.  M17-S1 is associated with CEN
16, an O9--B2 star. The larger bowshock,   M17-S2, is driven by CEN
18, an earlier-type star (O7--O8). 
Both
bowshocks were 
detected at 10.5 and 20 \um\ by \citet{MN01}. These observations did
not resolve the arc shapes of the 
bowshocks, and \citet{MN01} attributed the excess
IR emission to circumstellar disks and classified CEN 16 and IRS 9
(the star visible just to the right of the arrowhead in Figure
\ref{M17}) as massive protostars.

 M17-S3 lies outside of the field analyzed by \citet{HHC97},
and the 
driving star does
not appear in any catalog of the region. It is not found in any GLIMPSE
sourcelist,
because the bright, spatially variable diffuse background prevented
the automatic
extraction of the point source. We measured the flux this source
manually using a 5\arcsec\ aperture. Using the spectral energy distribution
(SED) fitting tool of \citet{fitter}, we
fit the IR fluxes of this star (Table 1) with \citet{Kurucz} stellar
atmosphere models. Following the method described by \citet{CW08}, we scale the
models to the 1.6-kpc distance of M17 and estimate a spectral type of
O7 V for the star. Carrying out the same analysis on 7 other O stars
in M17 with independently known spectral types \citep{MSP08}, we
estimate that our 
spectral typing is accurate to within 2 subclasses. 

\subsection{RCW49}

\begin{figure}[ht]
\begin{onecolumn}
\epsscale{0.8}
\plotone{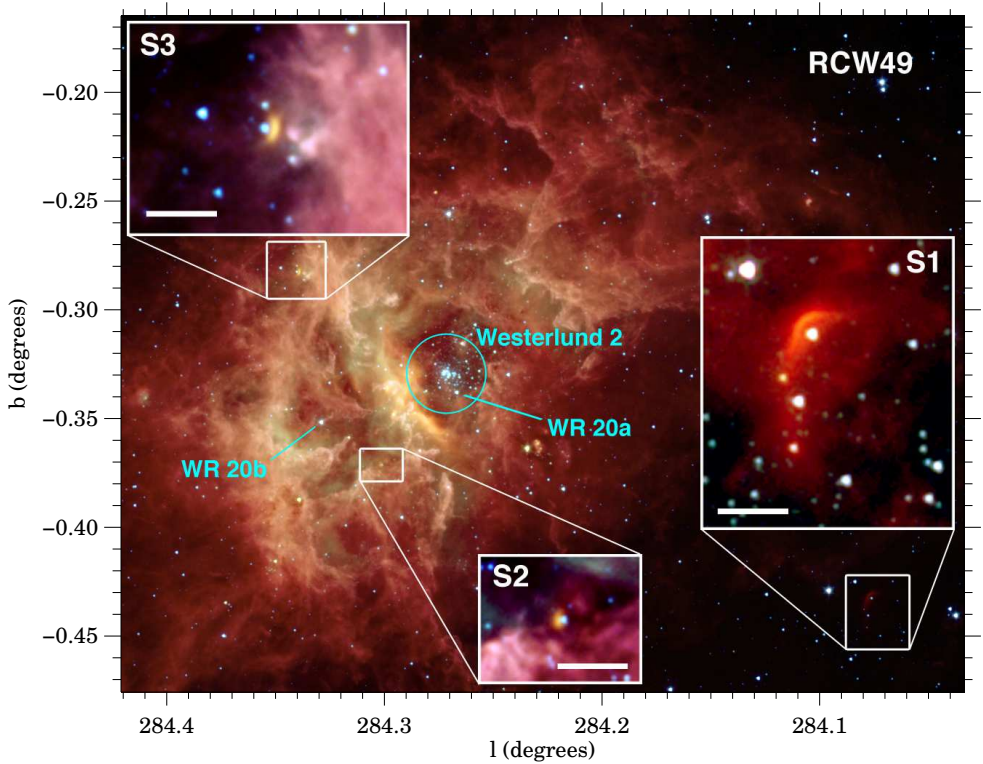}
\caption{GLIMPSE full-color image of RCW49 ({\it blue:} [3.6], {\it
    green:} [4.5], {\it orange:} [5.8], {\it red:} [8.0]). The
  bowshocks  RCW49-S1, -S2, and -S3, are enlarged in 3 separate insets
  (scalebars are $30\arcsec \approx 0.6$ pc at 4.2 kpc). 
  Three energy sources that could drive
  large-scale interstellar flows are also indicated: The Westerlund 2
  cluster (circled), and the Wolf-Rayet stars WR 20a and WR 20b.
\label{RCW49}}
\end{onecolumn}
\end{figure}

RCW49 presents a more complicated morphology than M17.
\citet{chu04} discussed the structure and spectrum of
the diffuse emission. Like M17, RCW49 is filled
with X-ray gas \citep{T05}. The interstellar structures are dominated by two
large cavities. The first, blown out to the West, contains the
massive young cluster 
Westerlund 2, and the second is an enclosed bubble around the
Wolf-Rayet star WR 20b (Figure \ref{RCW49}). 
The distance to RCW49
remains disputed, with recent estimates placing the region as close as
2.8 kpc \citep{JA07} 
or as far as 8.0 kpc \citep{GR07}, 
with a kinematic distance estimate of  $6\pm1$ kpc  \citep{dame07}. 
We consider distances of both 4.2 kpc \citep{chu04} and 6 kpc in our analysis. 

We have found three bowshocks associated with RCW49, shown in three separate
insets in Figure \ref{RCW49}.  RCW49-S1 is unique among our sample,
because it lies relatively far from the \hii\ region, 
but we believe it is associated with RCW49
for the following
reasons: (1) 
The RCW49
distance is consistent with the luminosity of the 
driving star being an O star (see below);
and (2) the bowshock points (approximately)
toward the central 
cluster, Westerlund 2. Because it is far from any other bright IR
source, RCW49-S1 appears as a point source in both the {\it Midcourse
  Space Experiment} \citep[{\it MSX};][]{MSX} and {\it Infrared
    Astronomical Satellite} \citep[{\it IRAS};][]{IRAS} point source
  catalogs, and these fluxes are also given in Table 2.

RCW49-S2 is oriented away from Westerlund 2
and appears to be influenced primarily by the
nearby WR 20b.  RCW49-S3 is oriented in the general direction
of Westerlund 2. None of the RCW49 bowshocks
points directly back toward the central cluster, while all three M17
bowshocks do. Perhaps the 
expanding bubbles driven by 
Westerlund 2 and the Wolf-Rayet stars are interacting turbulently,
producing non-radial components to the
flows. It is also possible that the bowshock driving stars have large
orbital motions relative to the dynamic interstellar medium.

The three bowshock driving stars in RCW49 are of previously
undetermined spectral type, so we again estimate the spectral type
by fitting model SEDs to the broadband fluxes of Table 1,  scaled to both the
4.2 kpc and 6 kpc distances. The results are given in Table 3.
\begin{deluxetable}{lccccccc}[ht]
\tablecaption{Bowshock Standoff Distances and Estimated Stellar Wind Properties\label{tbl-3}}
\tablewidth{0pt}
\tablehead{
\colhead{ID \#} & \colhead{Spectral} 
 & \colhead{$\dot{M}_{w,-6}v_{w,8}$\tablenotemark{b}}
 & \colhead{$d_w\cos i$} 
 & \colhead{$d_{cl}\cos i$} & \colhead{$v_{0}n_{0,3}^{1/2} (\cos{i})^{-1}$} \\
\colhead{} & \colhead{type\tablenotemark{a}} & \colhead{}
& \colhead{(pc)}& \colhead{(pc)} & \colhead{(km s$^{-1}$ )} }
\startdata
\multicolumn{5}{l}{M17 Distance $=1.6$ kpc} \\
\tableline
M17-S1 & O9--B2 V & $\leq 0.2$ &0.03 &   1.4 & $\le 22$ \\ 
M17-S2 & O7--O8 V &  0.5--2.5 & 0.11  &   0.9 & 10--22 \\
M17-S3 & O7 V &  0.5--2.5 & 0.05 &   1.7 & 21--47 \\
\tableline
\multicolumn{5}{l}{RCW49 Distance $=4.2$ kpc} \\
\tableline
RCW49-S1 & O6 V &  $\sim{1.3}$ &  0.16 & 16.2 & 11 \\
RCW49-S2 & O9 V & $\sim{0.2}$ &  0.04 & 2.46 & 17 \\
RCW49-S3 & O5 V or O9 III & 2.5--3.2 &  0.096  & 6.13 & 25--28 \\
\tableline
\multicolumn{5}{l}{RCW49 Distance $=6.1$ kpc} \\
\tableline
RCW49-S1 & O5 III &  $\sim{16}$  & 0.23 & 23.2 & 26 \\
RCW49-S2 & O6 V & $\sim{1.3}$ & 0.06 & 3.52 & 28 \\
RCW49-S3  & O3 V or O6.5 III &$> 3.2$ &  0.14 & 8.8 & $> 19$ \\
\enddata

\tablenotetext{a}{Spectral types for the stars driving M17-S1 and -S2
are taken from CEN and \citet{HHC97}. All others were estimated by
fitting Kurucz (1993) stellar atmosphere models to the broadband IR
fluxes (Table 1) and scaling to the distance of M17 or
RCW49. Spectral types given for RCW49 are highly uncertain due to the
disputed distance to that region.}
\tablenotetext{b}{Estimates of stellar mass-loss rates
  $\dot{M}_{w,-6}=\dot{M_w}/(10^{-6}$ 
  M$_{\sun}$ yr$^{-1}$) and wind velocities $v_{w,8}=v_{w}/(10^{8}$ cm
  s$^{-1})$ are based upon
  \citet{VdKL01} and \citet{FMP06}.} 

\end{deluxetable}
All three stars are plausibly O stars. Assuming 4.2 kpc, the driving star of
 RCW49-S1 is fit as O6 V,  RCW49-S2 as O 9 V, and  RCW49-S3 as
O5 V (or O9 III).   
If we increase the distance to 6 kpc, the fits become O 5 III, O6 V,
and  O5 V (or O6.5 III), respectively  \citep[see][for an explanation of the 
degeneracy between luminosity classes]{CW08}.
The spectral types at 6 kpc seem improbably luminous. 
Highly luminous and windy stars
dominate the dynamics of their local ISM.
We do not observe IR bowshocks around any
of the earliest-type stars in either RCW49 or M17, because they have
blown large, evacuated cavities in the centers of the \hii\
regions. Little or no ambient material remains close to the stars to produce
a bowshock.

Apart from the Wolf-Rayet systems, one of the
earliest stars in RCW49 
is G284.2642-00.3156. Using optical spectroscopy, \citet{BU05}
classified this star as O4 V(f)  
and derived a spectrophotometric distance of $3.2\pm 0.5$
kpc. Using IR SED fitting, we confirm that the luminosity of this star
is consistent with an O4 V star
at 3.2 kpc. Yet this distance falls on the low end of the range of published
distance 
estimates for RCW49, and we do not adopt it here. 
Like CEN 1 in M17, G284.2642-00.3156 may be an unresolved, equal-mass O4--O4 
binary. Doubling the luminosity moves the spectrophotometric distance from 3.2
to 4.5 kpc, in agreement with a distance of 4.2 kpc but inconsistent with 6 kpc.

\section{Bowshock Properties}
 
The standoff distance $d_w$ of a bowshock from its driving star is
the point where the momentum flux of the stellar wind balances the momentum
flux of the 
ambient medium: $n_{w}v_{w}^{2}=n_{0}v_{0}^{2}$. Following \citet{vBu88}, we
normalize the stellar wind mass loss rate,
$\dot{M}_{w,-6}=10^{-6}$ M$_{\sun}$ yr$^{-1}$, stellar wind velocity
$v_{w,8}=10^{8}$ cm s$^{-1}$, ambient hydrogen particle density of
$n_{0,3}=10^{3}$ cm$^{-3}$, and use a mean ISM gas mass per hydrogen atom
$\mu =2.36 \times 10^{-24}$ g. Assuming a spherically symmetric stellar wind
with a mass-loss rate given by $\dot{M}=4\pi
d_w^{2}\mu n_{w}v_{w}$, the velocity of the star 
with respect to the ambient ISM can be written as  
\begin{equation}
 v_{0}= 1.5\left(\frac{d_{w}}{{\rm pc}}\right)^{-1} (\dot{M}_{w,-6}v_{w,8})^{1/2}n_{0,3}^{-1/2}~[{\rm km~s^{-1}}], 
\end{equation} 
where $v_{w}$ is the terminal velocity of the stellar wind.
Values of
$v_{0}n_{0,3}^{1/2}$ and $d_w$ for each bowshock 
are presented in Table 3. The standoff distance can be measured only
as $d_w\cos{i}$ on the sky,
where $i$ is the inclination
 (or viewing angle) made by the line connecting the star with the apex
 of the bowshock 
against the plane of the sky 
($i=0$ if the line joining the star to the bowshock apex lies in the
plane of the 
sky). Because a bowshock oriented at high $i$ will not produce an arc
morphology, it is likely that $i \la 45\degree$, and hence $\cos{i}$
will not differ greatly from unity in our measurements.\footnote{The
  average value of $\cos{i}$ for $0\le i \le 45\degree$ is 0.9.} The 
distance from each bowshock to the likely source of the large-scale ISM flow,
measured on the sky as $d_{cl} \cos{i}$, are also presented for
reference in Table 3.

The mass-loss rates and stellar wind velocities in Equation 1 suffer from
high dispersion as a function of spectral type \citep{FMP06}, a factor
of 2 or even greater, and this is compounded by a comparable level of
uncertainty in the spectral types. The uncertainty on our measurements of $d_w$
ranges from ${\sim}20\%$ for the largest bowshocks (M17-S2 and
RCW49-S1) to ${\sim 40\%}$ 
for the smaller bowshocks that are barely resolved by IRAC.
 We estimate that
$v_0n_0^{1/2}$ is uncertain by a factor of 2 in M17 and up to a factor
of 3 in RCW49, where the spectral types of the driving stars are less
constrained. These uncertainties, reflected in the range of values
for $v_0n_0^{1/2}$ in Table 3, are dominated by the uncertainty in the
stellar wind properties.

We have neglected the effects of turbulent pressure in our calculation
of the momentum flux balance of the bowshocks. This is a potentially
significant contributor to the total ISM pressure held off by the
bowshocks. The effect of turbulent pressure would be to systematically
decrease the
standoff distance $d_w$, causing us to overestimate $v_0$.

In reality, for the fast winds of early-type stars, the observed
bowshock is displaced from the standoff distance by a
significant amount. This happens because the cooling timescale of the
shocked stellar wind is very long. The result is a thick layer of hot
gas intervening between the wind terminal shock at the standoff
distance and the thin, dense layer of interstellar gas and dust
forming the observed bowshock. The numerical simulations of \citet{CK98} predict
that the bowshock should be located at twice the standoff distance
from the driving star. In this case, our derived values of
$v_0n_0^{1/2}$ in Table 3 would be underestimated by the same factor
of 2. This systematic correction is comparable to the intrinsic
uncertainties in our estimates of $v_0n_0^{1/2}$, and it partially
compensates for the
effects of neglecting turbulent pressure in the ambient ISM. Therefore, the 
assumption that the observed distance of the bowshocks from the
driving stars corresponds to the standoff distance $d_w$ should not
have a large impact on our results, and we find that the cautious
application of Equation 1 
yields reasonable results.

Orbital velocities of O stars in massive clusters are typically ${<
  10}$ \kms. 
The expansion speed
of ionized gas in \hii\ regions 
is generally comparable to the sound speed of ${\sim}10$ \kms, and
this appears to be true 
in M17 \citep{EP07}. Most of
the bowshocks (the exception being  RCW49-S1) are apparently
located within the ionized gas of the radio \hii\ regions.   
The likely explanation for the bowshock emission in the IR is that
dust in the \hii\ regions \citep{MSP07} is swept-up by the bowshocks.
The observed average electron density in the Northern bar of the M17
\hii\ region is ${\sim}10^3$ cm$^{-3}$ \citep{FCM84}, so
the values of $v_0n_{0,3}^{1/2}$ listed in Table 3 are likely to be
close to the actual relative velocities of the stars and the ISM for
most of the bowshocks. 


The values of $v_{0}n_{0,3}^{1/2}$ calculated for
 M17-S1 and -S2
are in good agreement.
M17-S3,
however, presents a different picture. 
 M17-S3 appears to be associated with a ``teardrop'' structure (Figure 1) in the
 photodissociation region
(PDR), 
near the ionization front. The ambient density
surrounding this star could be significantly higher than the density
within the \hii\ region. If $v_0$ for this
bowshock is comparable to that of the other 2 bowshocks in M17, then
$n_{0,3} \sim 2.25$ cm$^{-3}$ in this location. This value agrees with
the measurements of electron density in the dense clumps of ionized
gas in M17 \citep{FCM84}. 
The unusual morphology of the 
diffuse IR emission associated with  M17-S3 suggests that the star
may have
recently emerged from an evaporating globule on the edge of the
PDR; a larger,
more evolved analog of the nearby pillar structure seen in Figure 1. 

In RCW49,
all 3 bowshocks are found in very different locations, but we note that 2 of
the bowshocks appear to be similar in size, color, and ambient environment
(Figure 2). The largest
bowshock,  
RCW49-S1, is different, since it is located
 relatively far (16.2 pc at 4.2 kpc) 
from Westerlund 2, outside the \hii\ region.
The presence of  RCW49-S1, along with its orientation, indicates
that RCW49 vents a large-scale flow of 
gas through the cavity opening to the West.
This flow likely originates in the combined winds of the Westerlund
2 cluster and thus should be much more diffuse than the \hii\ region
gas \citep{T03,T05}.  Assuming a density of 1 cm$^{-3}$ in the flow
from Westerlund 2, the standoff distance of  RCW49-S1 at 4.2 kpc
gives a flow velocity of ${\sim}350$ \kms. Such a high value of $v_0$
is reasonable, given that the gas must move supersonically relative to
the star to produce a shock, and the sound speed in the hot, rarefied
gas of the flow 
streaming away from the \hii\ region is ${\sim}100$ \kms.

\begin{figure}[ht]

\epsscale{0.5}
\plotone{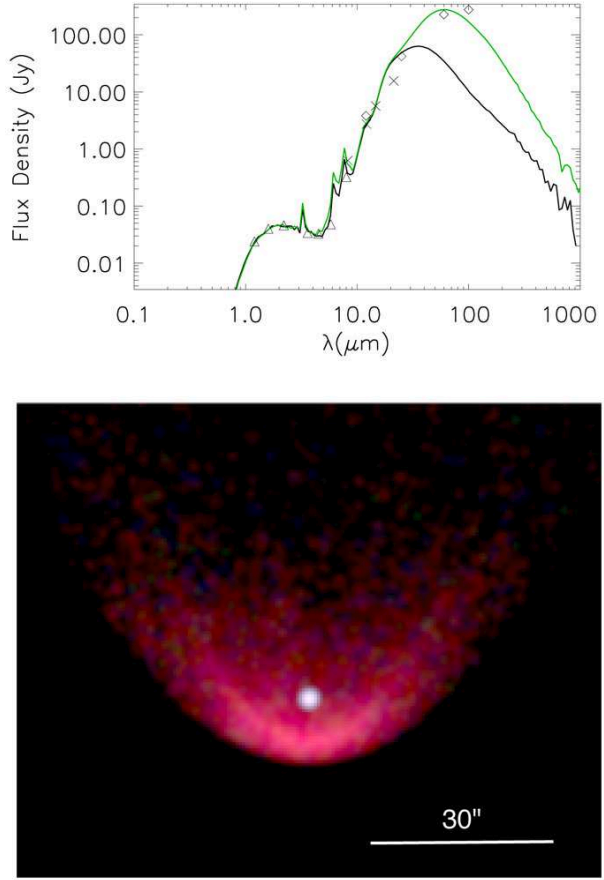}
\caption{Radiation transfer model of RCW49-S1. {\it Top:} Model SEDs
 plotted with the fluxes of the bowshock and driving star from Tables
 1 \& 2 ({\it triangles:} 2MASS and GLIMPSE; {\it crosses: MSX}; {\it
   diamonds: IRAS}). The {\it green} curve includes low-density
 material in a shell 2--3 pc from the star in order to match the {\it
   IRAS} 60 and 100 \um\ fluxes. {\it Bottom:} Image of the model 
 bowshock at GLIMPSE wavelengths
 (compare to Fig.\ 2).
\label{model}}

\end{figure}
Because RCW49-S1 was detected by {\it MSX} and {\it IRAS} in addition
to GLIMPSE, we can construct the SED of the bowshock from 4.5 \um\ to
100 \um\ (Table 2). We computed SED models of RCW49-S1 using a 3-D
radiative equilibrium code \citep{WI} 
modified to include very small grain (VSG) and PAH emission
\citep{KW08}. 
We used the canonical mass fraction for VSG/PAH grains of 5\%
\citep{DL07}. 
We modeled the bowshock geometry as a paraboloid with the apex offset
by $d_w=0.16$ pc  from the O6 V star (assuming $i=0$ and the 4.2 kpc
distance).  For models that reproduced the observed images, the SED
shape was insensitive both to the thickness of the shock and to the
radial density profile of the 
dust (with the total mass scaled to match the observed SED).
The black line in Figure \ref{model} shows the SED for a model with a
density varying as $r^{-2}$ from the 0.16 pc standoff distance out to
a radius of 1.5 pc.  The observed SED shortward of 30 \um\ is
well-matched by the model. 
For radial density exponents from $-2$ to 0, the mass of this material
ranges from ${\sim}0.5$ M$_{\odot}$ to 2 M$_{\odot}$, respectively, assuming a
dust-to-gas mass ratio of 0.01.  The optical depth in all models is
low, with $A_V < 0.02$ within 1 pc. To match the {\it IRAS} data at 60
and 100 \um\ requires low-density material farther from the star.  The
{\it green} line in Figure \ref{model} is a model including a shell
2--3 pc from the star.  This material added mid-IR PAH emission, so we
lowered the VSG/PAH mass fraction to 3\% to continue to match the
mid-IR SED, and the image still matches the data well. The $A_V$
through the
bowshock in this model is 0.25, most of it due to the outer shell.
Along the line of sight to the star, $A_V = 10$, so the
bowshock and shell contribute a negligible fraction of the
interstellar extinction.
These models show that dust distributed in a bowshock geometry matches
both the images and the SED reasonably well. The IR
emission from the bowshock can be explained by reprocessed stellar
radiation without any additional dust heating by the shock.

The structure of RCW-S1 is probably significantly different from that
of the other 5 bowshocks in our sample, because it is located in a
distinct interstellar environment. Because of the high temperature and
low density in the
flow outside the \hii\ region, the shocked interstellar gas cannot
cool quickly enough to form a dense layer behind the bowshock,
as likely happens in the other 5 cases. The shock is approximately
adiabatic, remaining very hot and only moderately compressed (by a factor
of ${\sim}4$) as it flows past the star, forming a relatively thick layer.
Hence, while RCW-S1 is the only bowshock in our sample
observed at enough 
different mid-IR wavelengths to allow us to create a meaningful model
of the emission, it may not be appropriate to
draw strong conclusions about the remaining bowshocks based upon this model.





\section{Summary}

We have observed 6 prominent IR bowshocks in M17 and
RCW49. These objects appear to be produced by the winds of
individual O
stars colliding with large-scale interstellar gas flows
in the \hii\ regions.
One bowshock,  M17-S3, may be the
leading edge of an evaporating globule containing a newly-formed and
previously undiscovered O star in the
well-studied M17 region. 
All three bowshocks associated with RCW49
lead us to identify new candidate O stars. Our stellar classifications
also suggest 
that the true distance to RCW49 is less than the kinematic distance of
6 kpc.

The bowshocks are bright at IR
wavelengths due to emission from dust swept up from the ambient
ISM and heated by radiation from the bowshock driving stars.
As \citet{AG08} note, IR excess emission from a bowshock could be
attributed to the presence of a circumstellar disk, particularly when
the bowshock 
morphology is not spatially well-resolved. This can be a pitfall for
observational studies of accreting massive protostars.
 

The collective
winds of the most luminous stars in young, massive clusters produce
overlapping large-scale flows that 
hollow out thermally hot cavities in the parent molecular cloud \citep{T03}.
The largest bowshock presented here,  RCW49-S1, is evidence that the
combined winds of the ionizing stars in Westerlund 2 have escaped the \hii\
region, creating a flow of hot gas moving at a few $10^2$ \kms\ that
extends at least 16 pc away from RCW49. 

The driving stars of the other 5 bowshocks
are surrounded by ionized gas and dust of their natal \hii\
regions, where the density of the ambient medium ($n_0\sim 10^3$ cm$^{-3}$)
is sufficiently high to produce the observed bowshocks with a relative
velocity of only 10--20 \kms.
The winds of
the bowshock driving stars do not directly encounter the ${>}2000$
\kms\ winds from the most massive stars in the cluster. Instead, the
bowshocks are shaped by the expansion of the ionized 
gas in the \hii\ regions relative to
 the orbital motions of the stars.

Eventually, supernova explosions
will produce high velocity shock waves that heat and disperse the original
gas cloud. In star forming regions like M17 and RCW49 that have not yet been
disrupted by supernovae, IR bowshocks serve as interstellar ``weather vanes,''
indicating the speed and direction of large-scale gas flows at points within
and around giant \hii\ regions. 

\acknowledgments

We thank the anonymous referee for incisive and very useful suggestions
that helped us improve this work. We are grateful to Joseph
Cassinelli, Eric 
Pellegrini, John Raymond, Ellen Zweibel, Heidi Gneiser, and Don Cox
for useful conversations while preparing this paper. 
M. S. P.  thanks the members of the ``dissertator club,'' Kathryn
Devine, K. Tabetha Hole, and Nicholas Murphy, for their helpful comments.
This work was supported by NSF grant AST-030368 (E. B. C.) and NASA/JPL
Contracts 1282620 and 1298148. Additional support was provided by the
NASA Theory Program (NNG05GH35G; B. A. W.). R. I. acknowledges support
from a Spitzer Fellowship at the time that these data were analyzed,
and from JPL RSA1275467.






\clearpage







\end{document}